\documentstyle[prb,aps]{revtex}

\begin{document}

 \draft

 \title{Anisotropic optical response of the diamond (111)-2$\times$1
surface} 

 \author{Cecilia Noguez$^{(a,b)}$ and Sergio E. Ulloa$^{(a)}$} 

 \address{$^{(a)}$ Department of Physics and Astronomy, and Condensed
Matter and Surface Sciences Program, Ohio University, Athens, Ohio
45701--2979. \\  $^{(b)}$ Instituto de F\'{\i}sica, Universidad
Nacional Aut\'onoma de M\'exico, Apdo. Postal 20-364, M\'exico
D.~F. 01000, M\'exico} 

 \date{26 November 1995}

 \twocolumn

 \maketitle

 \begin{abstract}

The optical properties of the 2$\times$1 reconstruction of the diamond
(111) surface are investigated. The electronic structure and optical
properties of the surface are studied using a microscopic tight-binding
approach. We calculate the dielectric response describing the surface
region and investigate the origin of the electronic transitions
involving surface and bulk states. A large anisotropy in the surface
dielectric response appears as a consequence of the asymmetric
reconstruction on the surface plane, which gives rise to the zigzag
Pandey chains. The results are presented in terms of the reflectance
anisotropy and electron energy loss spectra.  While our results are in
good agreement with available experimental data, additional experiments
are proposed in order to unambiguously determine the surface electronic
structure of this interesting surface. 

 \end{abstract}

 \pacs{PACS numbers: 78.66.Db, 73.20.At, 78.66.-w, 73.20.-r}  

 \narrowtext

\section{Introduction}

Apart from being of fundamental interest, the characterization of the 
low-index diamond surface is very important from a technological point
of view. The fast development of chemical vapor deposition techniques
has increased the demand for a better understanding of the ground state
and excited properties of these surfaces.\cite{celii} Indeed, much
experimental and theoretical attention has been paid to the
characterization of the geometrical structure, vibrational modes and
electronic properties of these surfaces, with interesting and sometimes
controversial results. In this work, we are interested in
characterizing the {\em optical} response of the (111) diamond surface,
and investigating how these properties are related to the structural
reconstruction and its accompanying electronic structure.  

Our interest on this particular surface includes concerns on
discrepancies between the present experimental 
\cite{himpsel,pate,pepper1,lurie,morar,kubiak} and theoretical 
results.\cite{vanderbilt,iarlori,alfonso,davidson}  Experimentally, a
great deal of the surface electronic structure is well known through
angle resolved photoemission spectroscopy (ARPES),\cite{himpsel,pate}
soft-x-rays absorption,\cite{morar} inverse photoemission,\cite{kubiak}
and electron energy loss spectroscopy (EELS).\cite{pepper1,lurie} The
photon-induced measurements show a variety of occupied
\cite{himpsel,pate} and unoccupied \cite{morar,kubiak} surface states
lying in the fundamental gap. However, the complete description of
these surface states has been difficult, since only their dispersion
along a few of the main directions of the surface unit cell have been
measured. On the other hand, EELS measurements \cite{pepper1} show a
prominent broad feature at about 2.1~eV, which is attributed to
transitions from occupied to unoccupied surface states.  Since EELS
experiments measure a transition energy which is generally smaller than
the difference between occupied and unoccupied states, and a relatively
large uncertainty ($\pm 0.6$~eV) accompanied this particular EELS
experiment, a direct comparison with other results has not been
possible.

Several theoretical studies have been done to elucidate the structural
and electronic properties of the C(111)-2$\times$1 surface. Both, {\em
ab initio} \cite{vanderbilt,iarlori,alfonso} and semi-empirical
\cite{davidson} theoretical approaches have been employed, yielding
some differences among them and with experimental results. Some of
these differences arise from the methodology employed.  For example,
Iarlori and co-workers \cite{iarlori} employed a LDA formalism using a
plane wave basis, while Vanderbilt and Louie,\cite{vanderbilt} and
Alfonso and co-workers \cite{alfonso} used the LDA formalism based on a
set of localized orbitals. In the former work the energy gaps are
underestimated,\cite{iarlori} as is common in this kind of
approximation, and a direct comparison to experimental results is
difficult.  The latter theoretical works compare well among
them,\cite{vanderbilt,alfonso} although a systematic shift of about
1~eV is found when the surface states are compared with those measured
experimentally.\cite{himpsel,pate,morar,kubiak}  On the other hand, the
semi-empirical tight-binding approach of Davidson and Pickett 
\cite{davidson} compares well with the {\em ab initio} results
described above, except for an extra shift of the surface states by
about 0.8~eV.  Since the surface states determine the location of the
Fermi level, there is a large discrepancy among different theoretical
works as to the relative position of the Fermi level and the top of the
valence band that goes from --1.3 to 2~eV.\@  On the other hand, in all
theoretical and experimental results there is general good agreement on
the energy gap between empty and full surface states.   From these
considerations, one can then conclude that the calculated electronic
structures alone are not able to uniquely determine the nature of the
transitions observed in EELS,\cite{pepper1,lurie} resulting in
controversial interpretations of the available experimental optical
data.\cite{himpsel,pate,morar,kubiak}  The evaluation of the surface
dielectric response function for this system, and its analysis in terms
of the associated electronic level structure gives further insights
into this problem, as we discuss below.

In the present work, and in close connection with the general
description of the optical properties of the C(111)-2$\times$1 surface,
we investigate in detail the origin of the electronic transitions
related to the surface reconstruction.  Our calculations employ a
semi-empirical tight-binding approach that has been used previously to
study the optical properties of Si(111), (110) and
(100),\cite{selloni,mochan,shkrebtii,noguez} and C(001)
\cite{gavrilenko} surfaces. Our tight-binding formalism is similar to
the one used by Davidson and Pickett,\cite{davidson} except that our
extended orbital basis allows perhaps a better description of the
conduction band due to its additional s$^*$ orbital, and our level
structure is in general better agreement with experimental findings and
other calculations.

Since we calculate the {\em surface} dielectric tensor, the results
presented here can be compared directly with those measured using
various optical spectroscopies. In particular, the differential
reflectance and reflectance anisotropy spectroscopies provide accurate
information about surface properties of metals \cite{tarriba} and
semiconductors.\cite{chiaradia}  This is very important since several
semiconductor surfaces show a metallic-like behavior due to the narrow
gap between occupied and unoccupied surface states that hamper the use
of electronic spectroscopies like ARPES, EELS and Scanning Tunnelling
Microscopy (STM). This indeed seems to be the case for the
C(111)-2$\times$1 surface, where a narrow gap has been found
theoretically along one of the main directions on the
surface,\cite{vanderbilt,iarlori,alfonso,davidson} while no direct
experimental evidence is found in the literature for this metallic
behavior. 
 This behavior is in comparison with the Si and Ge (111)-2$\times$1
reconstructed surfaces,\cite{northrup1,northrup2} where the degeneracy
of the surface states is broken by the buckling of the surface atoms,
as the theoretical and experimental description of the surface states
shows.   Notice, furthermore, that the optical spectroscopies mentioned
above have the advantage over other techniques of allowing {\em in
situ} real-time measurements, which provide the invaluable opportunity
of monitoring the chemical vapor deposition and molecular beam
epitaxial growth,\cite{turner} as well as the dynamics of the
chemisorption process.\cite{roy}  The results presented here then not
only provide answers to fundamental questions, but give important
information for applications, which we expect will motivate future work
in this direction. 

In Section II, we present a brief discussion of the structural model of
the surface and the methods used to calculate its electronic and
optical properties. In Section III, we discuss our results and compare
with the available data in the literature. The results for the optical
properties are presented in terms of the dielectric response of the
surface, and the calculated reflectance anisotropy and EELS spectra. 

\section {Models and method of calculation}

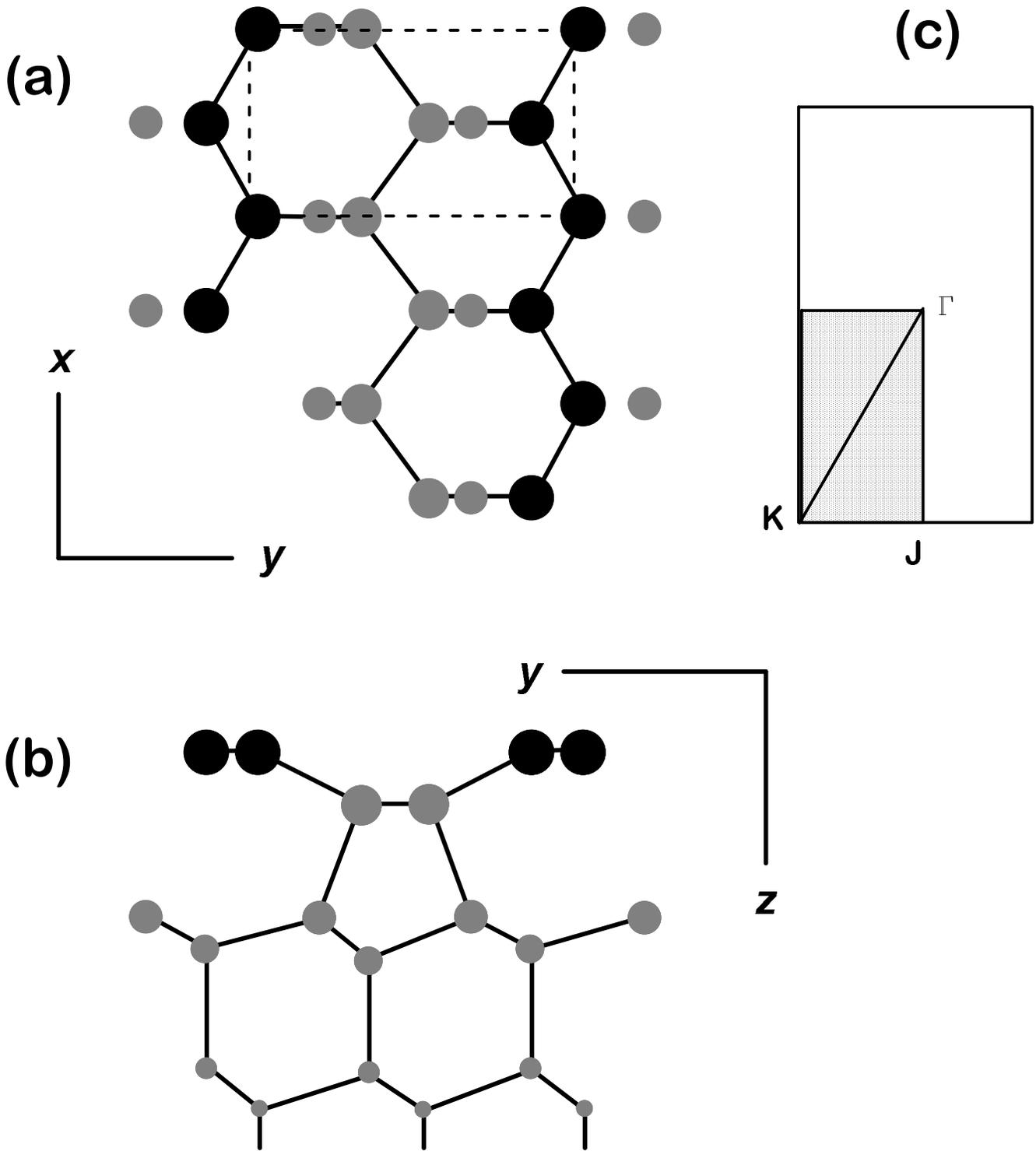
 \begin{figure}
 \caption {Atomic model of the C(111)-2$\times$1 surface. (a) Top view
with the three uppermost layers; dashed line corresponds to the surface
unit cell. (b) Side view with the six uppermost layers. The first-layer
atoms forming Pandey-like chains shown in black. (c) Surface Brillouin
Zone is shown; shadowed area corresponds to its irreducible part. The
main symmetry points are indicated.}
 \label{modelo}
 \end{figure}

The diamond (111)-2$\times$1 surface was modeled using a slab of 28 C
layers with inversion symmetry, yielding a free reconstructed surface
on each face of the slab. The thickness of the slab is large enough to
decouple the surface states at the top and bottom surfaces of the slab.
In Fig.\ \ref{modelo}, we show (a) the top view of the surface unit
cell that contains two C atoms per layer, (b) a side view with only the
six outermost layers of the slab and (c) the irreducible surface
Brillouin zone (SBZ). Periodic boundary conditions were employed
parallel to the surface of the slab to effectively model a
two-dimensional crystal system. The top (and bottom) layer of the slab,
shown in Fig.~\ref{modelo} with black circles, resemble the structure
reported by Pandey.\cite{pandey} In this Pandey chain model, the atoms
of the top layer form a zigzag chain along one of the main directions
on the surface plane (the $x$ axis in Fig.\ \ref{modelo}).  

The coordinates for the six outermost layers on each side were obtained
by Alfonso {\em et al}.,\cite{alfonso} using a first-principles density
functional based molecular dynamics technique due to Sankey and
co-workers.\cite{sankey}  (The remaining central layers have bulk
 geometry.)  The method has been employed successfully in  studying
covalent systems such as silicon and carbon.\cite{sankey,r20,r21} The
relaxed C(111)-2$\times$1 surface obtained with this method showed the
zigzag-like chains with no buckling on the surface layer, and with CC
bondlengths of about 1.44 \AA. The results of Ref.\ [\ref{alfonso}] are
in excellent agreement with previous self-consistent first-principles
calculations,\cite{iarlori,marcus} where they find unbuckled surface
chains with bondlengths equal to 1.47 and 1.44 \AA, respectively. The
reader is referred to Ref.\ [\ref{sankey}] for a comprehensive
description of this technique, and to Ref.\ [\ref{alfonso}] for a
detailed discussion of its applicability to diamond surfaces.  The use
of the fully relaxed slab coordinates guarantees that the optical
properties we calculate include all the subtle effects of
surface-induced strain and appropriate geometry.

To calculate the optical properties of the system, we generate the
electronic level structure of the slab using a well known parameterized
tight-binding approach with a sp$^3$s$^*$ orbital basis.\cite{vogl}
This basis provides a good description of the conduction band of cubic
materials. This approximation has been applied to calculate the optical
properties of a variety of silicon surfaces,
\cite{selloni,mochan,shkrebtii,noguez} and recently to the (001)
surface of diamond, \cite{gavrilenko} yielding good results. The
parameters for CC interactions are taken to be the same as those of
Ref.\ [\ref{vogl}] for the bulk, except for the on-site energy of the
p$_z$ orbitals of the surface atoms, E$_{\rm p}$.  This parameter is
set to be 2.3 eV smaller than the corresponding bulk parameter. This
change is assumed to be the likely result of additional orbital
confinement at the surface, and as we will see below, it yields a level
structure more attuned to experiments and other theoretical
calculations.  Moreover, the scaling factor of all tight-binding
parameters for this particular surface was taken as $( r/r_\circ)^7$,
where $r$ is the bondlenght of any two first-neighbors atoms and
$r_\circ =1.56$ \AA, is the bondlenght in bulk diamond.  These changes
to the original bulk parameters provide an excellent description of the
electronic structure, as compared to experimental measurements,
\cite{himpsel,pate,morar,kubiak} as we will show on Section III.   

The optical properties of the surface region are determined by its 
dielectric function. The imaginary part of the average slab
polarizability is related to the transition probability between slab
eigenstates induced by an external radiation field.\cite{bassani}
Within a single-particle scheme, this relation is expressed by  
 \begin{eqnarray}
{\rm Im} \, \alpha_{\rm slab}^{\alpha \alpha} (\omega)&=& \frac{\pi
e^2}{m^2 \omega^2 A d}  \sum_{\bf k} \sum_{v, c} | p^\alpha_{vc} ({\bf 
k})|^2 \cr 
& & \hskip.7in \times \delta(E_c({\bf k}) -E_v({\bf k}) - \hbar \omega),
\label{alfa} 
 \end{eqnarray}
where $ p^\alpha_{vc} ({\bf k})$ is the matrix element of the $\alpha$ 
component of the momentum operator between valence ($v$) and
conduction ($c$) states at the point {\bf k} of the SBZ, $2d$ is the
slab thickness, $m$ is the bare electronic mass, and $A$ is the area
of the surface unit cell. The real part of the average polarizability
can be computed via the Kramers-Kronig relations. The surface
dielectric tensor $\epsilon_{\rm surf}^{\alpha \alpha} (\omega) = 1 +
4 \pi \alpha_{\rm surf}^{\alpha \alpha} (\omega)$, is then calculated
from the average slab polarizability  \cite{delsole1}
 \begin{equation}
d \alpha_{\rm slab}^{\alpha \alpha} (\omega) = d_{\rm surf} \alpha_{\rm 
surf}^{\alpha \alpha} (\omega) + [d -d_{\rm surf}] \alpha_{\rm bulk}
(\omega)  \delta_{\alpha \alpha}. 
\label{df}
 \end{equation} Here $\alpha_{\rm bulk} (\omega) = [ \epsilon_{\rm
bulk} (\omega) -1 ] / 4 \pi $ is the bulk polarizability, and $d_{\rm
surf}$ is the depth of the surface region. Note that for cubic
materials (C, Si, and Ge, for example) the bulk dielectric function is
isotropic. The ``three-layer model'' of Drude \cite{drude} and McIntyre
and Aspnes \cite{mcintyre} adopted here is widely used in the analysis
of optical data, and assumes that the system consists of three
homogeneous regions: bulk, surface and vacuum, and the dielectric
response is treated accordingly.

The matrix elements of the momentum operator $p^\alpha_{vc} ({\bf k})$
of Eq.~(\ref{alfa}), were obtained in terms of the atomic-like orbital
basis using the commutation relation between the Hamiltonian and
position operator, $ {\bf p} = i(m/\hbar)\, [ H, {\bf r} ] $. Taking
advantage of the orthogonality and localization of the orbitals, only
the intra-atomic dipole matrix elements are retained. Then, only two
additional parameters to those of the tight-binding Hamiltonian were
needed in order to reproduce the bulk dielectric function. These
parameters are the so-called intra-atomic sp and s$^*$p dipoles, with
best fitted values of 0.18~\AA~and 0.7~\AA, respectively.  Notice that
these calculations neglect in principle excitonic \cite{delsolex} and
local field effects,\cite{mochanlf} although the fitting parameter
procedure compensates to some extent and yields very good agreement
with bulk optical measurements. For a detailed description of the
method the reader is referred to the pioneering work of Selloni {\em et
al}.\cite{delsole1} and the review by Del Sole.\cite{delsole2}

In the above discussion, we have seen that the atomic structure of the 
surface region is intimately related to the dielectric response through
its electronic structure, as given by Eq.~(\ref{alfa}). Experimentally,
it is known that the surface dielectric function can be extracted by
means of electronic and optical spectroscopies.  Measurements of the
reflectance anisotropy (RA) is one of these optical techniques which
consists of measuring the relative reflectance difference of two
orthogonal light polarizations on the surface plane, $x$ and $y$ for
example.  Although the sample penetration of light is in general a few
hundred times larger than the depth of the surface layer, the
contribution from the bulk region to the RA spectra is canceled since
the bulk optical properties of cubic materials are isotropic. 
Correspondingly, this technique is extremely sensitive to surface
features and electronic properties due to reconstructions or adsorption
events.

Theoretically, the reflectivity is related to the dielectric function
through the Fresnel formula,\cite{Jackson} which must however, be
modified due to the presence of the reconstructed surface
region.\cite{drude,mcintyre,barrera,schaich}  This correction yields
the following expression for the differential reflectance spectrum when
the light incidence is normal to the surface plane,\cite{delsole1} 
 \begin{equation} 
 \left( \frac{ \Delta R}{R_\circ} \right)^\alpha  = \frac{ 8 \pi \omega
d} {c}\, {\rm Im}\, \left[ \frac{ \alpha_{\rm slab}^{\alpha \alpha}
(\omega)}{ \epsilon_{\rm bulk} (\omega) -1} \right].  \label{dr}
 \end{equation} 
 Here, $\alpha$ is one of the orthogonal directions on the surface
plane, $\Delta R = R -R_\circ$ is the difference between the actual
reflection coefficient $R$ and the reflectivity $R_\circ$ given by the
Fresnel formula.  

The second experimental technique in which we are interested is the
electron energy loss spectroscopy (EELS). Here, an electron beam of a
given low-energy and momentum is scattered by the sample. The electron
beam induces polarizations on the surface region so that the  electrons
lose some of this energy before being scattered into the detector. The
process can be described well in terms of a dipolar scattering
theory,\cite{ibach} and provides a suitable description of vibrational
modes of surface atoms and molecules, as well as electron transitions
on the surface region.  In the present work, all of the electronic 
transitions in the surface region are due to the reconstruction of the
surface and not to adsorbates, although the work could be generalized
to include various adsorbate species as well.

The electron scattering probability $P({\bf q}_\|, \omega)$ for an
electron that loses a quantum of energy $\hbar \omega$, and transfers
a momentum $\hbar {\bf q}_\|$ in the direction of the surface plane is
given by \cite{ibach}
 \begin{equation}
P({\bf q}_\|, \omega) = \frac{2}{(e a_o \pi)^2} \frac{1}{\cos \varphi_i}
\frac{k'}{k} \frac{q_\parallel}{|q_\parallel^2 + q_\perp^2|^2} \,\,
{\rm Im}\, g({\bf q}_\parallel, \omega),
\label{pd}
 \end{equation} 
where {\bf k \rm and \bf k}$'$ are the wave-vectors of the incident and
scattered electrons, $\varphi_i$ is the angle of incidence and $\hbar
q_\perp = \hbar (k_z -k_z') $ is the momentum transfer in the direction
perpendicular to the surface plane. The above relation holds when the
energy loss and momentum transfer to the medium are small. Assuming
that the scattering occurs in the $yz$ plane, the loss function is
defined by
 \begin{equation}
{\rm Im} \, g(q_y,\omega) = {\rm Im} \,\left( {-2 \over 1 +
\epsilon_{\rm eff} (q_y, \omega)} \right), 
 \end{equation}  
 where $\epsilon_{\rm eff}(q_y, \omega)$ is the nonlocal effective
dielectric function of the system. In the limit $q_z d_{\rm surf} \ll
1$, when the momentum transfer to the medium in the perpendicular
direction to the surface plane is small, this effective dielectric
function becomes
 \begin{eqnarray} 
 \epsilon_{\rm eff} ( q_y, \omega ) &\approx& \epsilon_{\rm bulk}
(\omega) \cr 
 &+& q_y d_{\rm surf} \left[ \epsilon_{\rm surf}^{y y} (\omega) - 
\epsilon_{\rm bulk}^2 (\omega)/ \epsilon_{\rm surf}^{z z} (\omega) 
\right]. 
 \end{eqnarray}
 This theory has been applied successfully to explain the experimental
EELS spectra of the 2$\times$1 and 7$\times$7 reconstructions of the
Si(111) surface.\cite{noguez,delsole3}

In the following section we use our calculated surface dielectric
function to explain the main features of RA and EELS spectra of the
C(111)-2$\times$1 surface.  

\section{Results and discussion}

\subsection{Surface Band Structure}

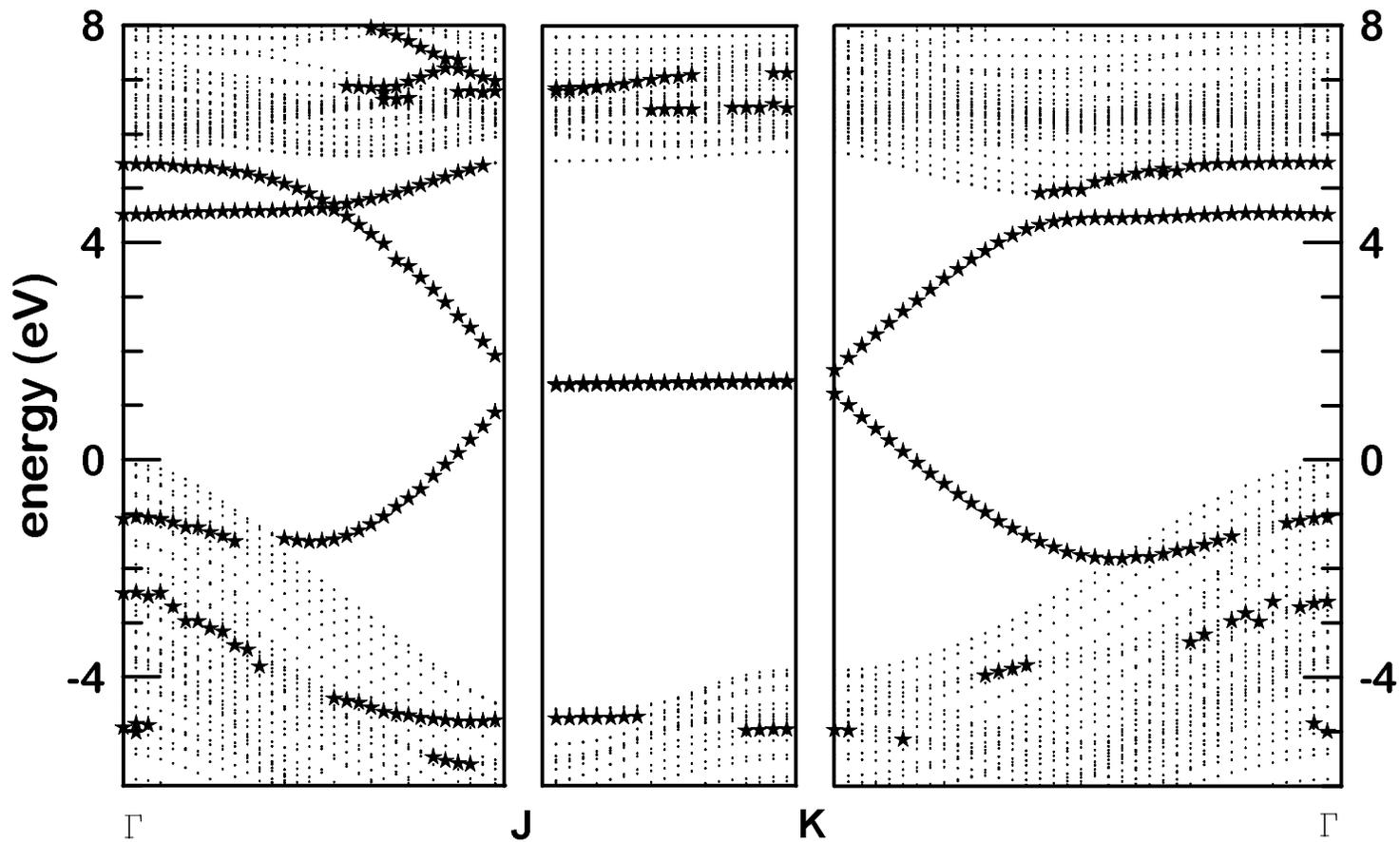
 \begin{figure} 
 \caption {Surface electronic structure along the main symmetry
directions of the surface unit cell. Dots correspond to the
projected bulk states, while stars represent surface states. Resonance
states embedded in projected bulk bands are also represented by stars.}
 \label{sbs}
 \end{figure}

The electronic band structure of the C(111)-2$\times$1 surface is
presented in Fig.~\ref{sbs}. The electronic structure is shown along
the main symmetry directions of the irreducible SBZ, from $\Gamma$  to
J ($x$ direction), from J to K ($y$ direction) and from K to $\Gamma$
(diagonal). The states associated to the surface reconstruction are
represented by stars, while dots correspond to the projected bulk
states. The top of the bulk valence band is set at 0~eV, and the
calculated Fermi level ($E_f$) is at about 1.5~eV (not indicated in
Fig.~\ref{sbs}), and coincident with the nearly-degenerate and flat
dispersion states along JK\@.  This result is in excellent agreement
with the reported experimental value of
1.5$\pm$0.2~eV.\cite{himpsel,pate,morar,kubiak}  In fact, the
calculated results presented here are in excellent general agreement
with experimental measurements \cite{himpsel,pate,morar,kubiak} and
also compare  well with those calculated previously using
first-principles \cite{vanderbilt,iarlori,alfonso} and parameterized
tight-binding \cite{davidson} approaches. 

The calculated surface band structure of Fig.~\ref{sbs} shows a large
gap of about 5.5~eV between the occupied and unoccupied  surface states
at the $\Gamma$ point. The occupied surface states liying within the
bulk valence band show a dispersion of $\sim$ 2.4~eV along the
$\Gamma$J direction (the Pandey chain axis direction), and have mainly
a p$_z$ character. The behavior of these states is similar to one
observed experimentally by Himpsel and co-workers \cite{himpsel} and
Pate and co-workers \cite{pate}, where a nearly flat filled surface
band is found from $\Gamma$ to about 0.5$\Gamma$J where there is a
minimum, and then rapidly disperses upward while approaching the J
point.  A similar behavior is found for these surface states along the
$\Gamma$K direction. Above the Fermi level there are two bands of
unoccupied states near 4.5 and 5.5~eV at the $\Gamma$ point.  These
states have a strong s and p$_z$ component corresponding to the
dangling bonds of the surface chain atoms and the backbonds with the
second and third layer atoms. The states at 4.5~eV show a nearly flat
band in the first half of the $\Gamma$J and $\Gamma$K directions, at
about the halfway point in both directions the band has a maximum and
then rapidly disperses downward approaching the J and K points. In
the direction perpendicular to the chain, JK, these empty surface
states and the occupied surface band become nearly degenerate and
show little dispersion of less than 0.1~eV. Notice that these two
states cross the Fermi level halfway through the JK direction. The
striking difference in dispersion of the surface bands along the two
main directions is a reflection of the Pandey-like chains formed on
the surface. The chains along the $\Gamma$J (or $x$) direction allow
for nearly free-like electronic motion (although the dispersion is
not parabolic), while the nearly vanishing overlap between the chains
along JK (or $y$) direction yields nearly flat surface bands and
reduces electronic hopping across zigzag chains.

Near $\sim$5.5~eV at the $\Gamma$ point begins a band of unoccupied
surface states mainly due to the surface dangling-bonds, and part to
second layer backbonds, corresponding approximately to those
calculated by Vanderbilt and Louie \cite{vanderbilt} and Alfonso and
co-workers, \cite{alfonso} and likely to be those observed by Kubiak
and Kolasinski.\cite{kubiak}  These states show a dispersion of about
1~eV with a minimum at about halfway the $\Gamma$J direction, where
they anticross the surface band associated to the backbond states
described above, and produce a hardly noticeable splitting at the
crossing.  On the other hand, these states show less dispersion
($\sim$0.5~eV) along the $\Gamma$K direction and never cross the
empty surface dangling bonds band.  These states have also been 
observed experimentally by Kubiak and Kolasinski \cite{kubiak} with a
weak intensity at an energy of about 5.8~eV from the top of the valence
band, for both the $\Gamma$J and $\Gamma$K directions. Finally, some
localized resonance-like occupied states are also found at $-2.5$~eV
and at $-4.8$~eV near the $\Gamma$ point (shown as stars within the
valence band in Fig.\ \ref{sbs}).  The former states are mainly due to
the subsurface chains with a strong p$_x$ component, and the latter one
have first- and second-layer backbond characteristics. The states at
$-2.5$~eV are similar to those reported by Vanderbilt and
Louie.\cite{vanderbilt}

Experimental photoemission results show occupied \cite{himpsel,pate}
and unoccupied \cite{morar,kubiak} surface states with a gap of nearly
5.1~eV at the $\Gamma$ point. The Fermi level is reported at about
1.5~eV above the top of the valence band, similar to our findings,
while the dispersion of the observed surface states is in very good
agreement with those calculated here, and those reported by Vanderbilt
and Louie.\cite{vanderbilt}  Moreover, since these surface states are
detectable only for p-polarization, they show a strong s and p$_z$
character, in agreement with our results. A resonance unoccupied state
has also been observed at about 6~eV from the top of the valence band
at the $\Gamma$ point.\cite{kubiak} This resonance state is much
weaker than the surface states lying in the fundamental gap, and it has
not been possible to fully investigate its orbital character, although
it could be the higher-energy surface state we find.

While the calculated surface bands in Fig.\ \ref{sbs} compare well with
the experimental results,\cite{himpsel,pate,morar,kubiak}  the
calculated LDA \cite{vanderbilt,alfonso} and tight-binding
\cite{davidson} results are rigidly shifted by +1 and +2 eV,
respectively. The discrepancies among the different approaches could
perhaps be partly attributed to many-body effects. For example, when
the exchange correlation effects are considered, the surface band is
shifted towards the top of the projected bulk valence
band.\cite{vanderbilt,iarlori}  Moreover, when dynamical effects are
taken into account within the GW approximation, the surface band moves
into the projected bulk valence band in the vicinity of the $\Gamma$
point,\cite{kress} also in agreement with experimental
results.\cite{himpsel,pate} However, there could be other sources of
error when we compare directly with experimental results, including the
precise experimental location of the Fermi level, as pointed out
before.\cite{vanderbilt}

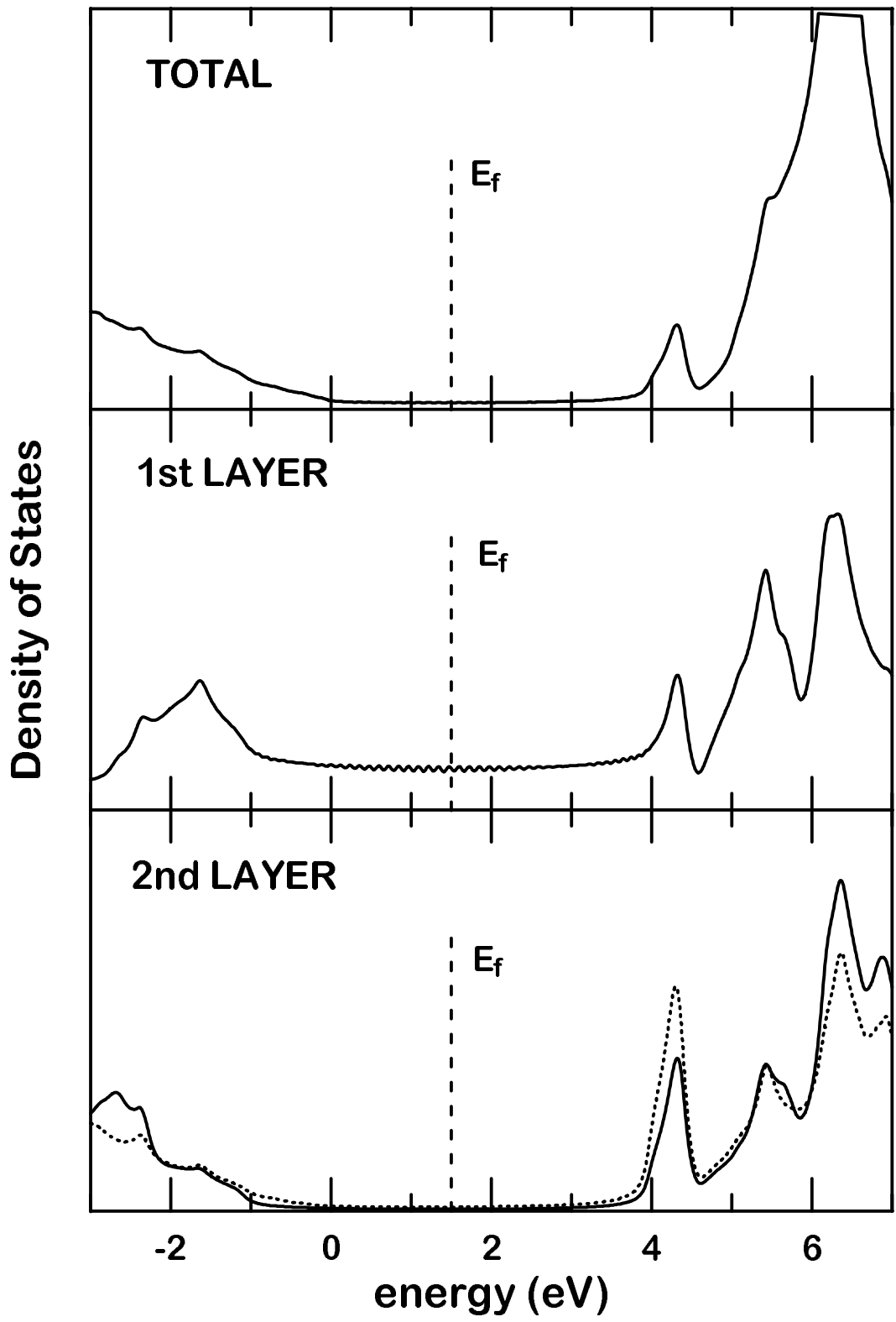
 \begin{figure} 
 \caption {Calculated density of states for (a) {\em TOTAL}, (b)
projected on first layer {\em 1st LAYER}, and (c) projected on {\em 2nd
LAYER} (solid) and {\em 3rd LAYER} (dotted line). Different vertical
scales used in each panel.}  
 \label{dos}
 \end{figure}

The total electronic density of states (DOS) of the slab, and the
projected density of states of the first two layers are shown in
Fig.~\ref{dos}. The DOS was calculated taking an average over 4900
points distributed homogeneously in the irreducible SBZ. We observe
within the fundamental bulk gap, between 0 and 5.5~eV, a non-zero
density of states coming mainly from the dangling bonds associated with
first layer atoms. This continuum of surface states is responsible for
the metallic-like behavior of the surface around the Fermi level, as we
will discuss in detail in the next section.  Notice that the DOS from 0
to 4~eV is nearly a constant, as expected for a 2D free-electron
system.  From Fig.~\ref{dos} is clear that the peak at about 4.5~eV
with a strong p$_z$ component has its origin in the dangling bond and
the first- and second-layer backbonds. The resonance states at about
$-2.5$ and $5.5$~eV are associated with the second layer chains and the
backbonds between the first layer and second layer atoms, respectively.
The pronounced peaks of the projected DOS in the first layer, at about
$-1.5$ and 4.5~eV are due to the lack of dispersion of the surface
bands on the first half of the $\Gamma$J and $\Gamma$K directions of
the SBZ (see Fig.\ \ref{sbs}). Here, we observe that the states in the
bulk gap are mainly localized in the first two layers, as one would
expect, with decreasing intensity into the slab.

Finally, before addressing the optical consequences of this level
structure, we should comment on our choice of parameters.  The
excellent agreement with experiments and {\em ab initio} electronic
calculations has been greatly enhanced by our use of the different
E$_{p_z}$ parameter at the surface atoms, as mentioned above, as well
as  to the fully-relaxed atomic positions for the reconstructed surface
of Ref.\ \ref{alfonso}.  Indeed, use of the {\em bulk} E$_{p_z}$
parameters for all surface atoms yields a level structure (not shown)
very similar to that of Davidson and Pickett.\cite{davidson}  In that
case, we obtain $E_f \approx 3.5$ eV above the valence-band top, while
the  filled surface dangling bond state remains $\approx 2.7$ eV below
$E_f$ (but now {\em above} the valence band).  Similarly, the gap
between surface state and conduction band bottom along the JK direction
is only $\approx 2.5$ eV, rather than the 4 eV gap shown in Fig.\ 2. 
This full set of results validates the choice of the physical parameter
E$_{p_z}$ at the surface.  Although a detailed fit to the experimental
results was not performed, it is clear (as one would expect on general
physical
 grounds) that the orbital localization at the surface affects the 
diagonal tight-binding parameters.  A full {\em ab initio}
determination of the various optical parameters, both in the bulk and
near the surface, together with the fully relaxed level structure will
be obviously desirable.  We are currently carrying out such project and
our results will be presented elsewhere.

\subsection{Surface Dielectric Properties} \label{optica}

The imaginary part of the average polarizability of the slab,
Eq.~(\ref{alfa}), was calculated using 4900 points distributed
homogeneously on the irreducible SBZ. The large number of points needed
is due to the small (large) size of the surface unit cell in real
(reciprocal) space and to the large sections of the SBZ with flat joint
density of states. The average over this large number of points is
necessary to give full and reliable convergence of the optical
properties for this particular surface. Electron transitions up to
20~eV were taken into account, so that after the Kramers-Kronig
transform the calculated real part is accurate up to about 10~eV.

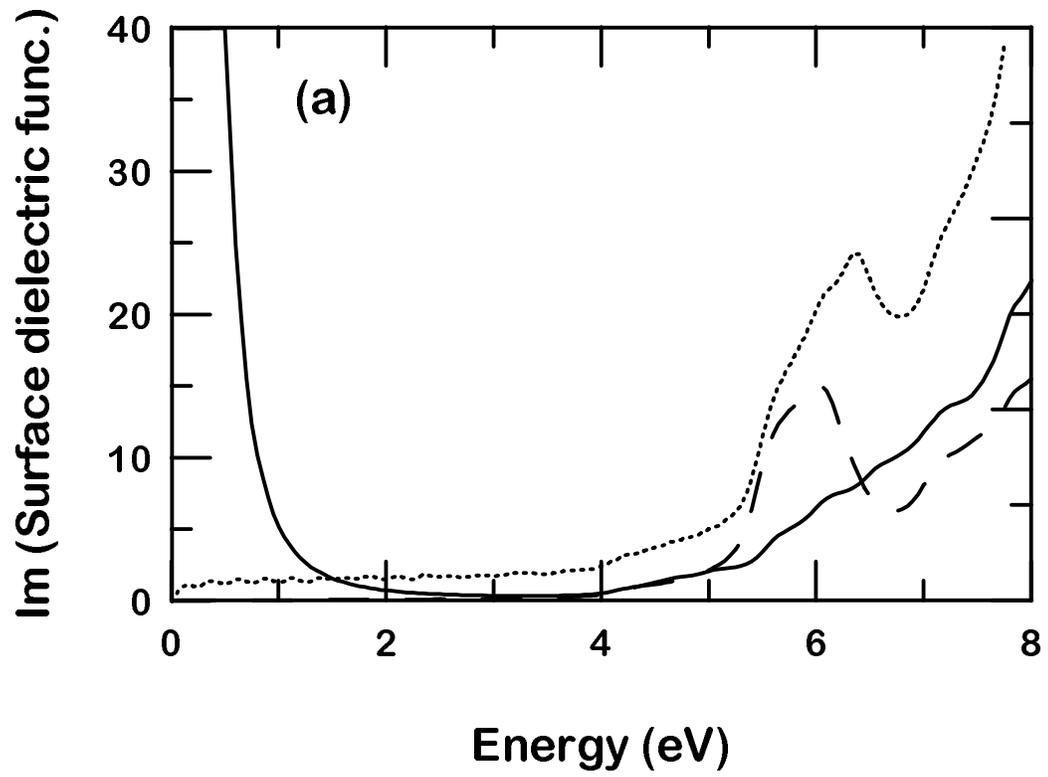
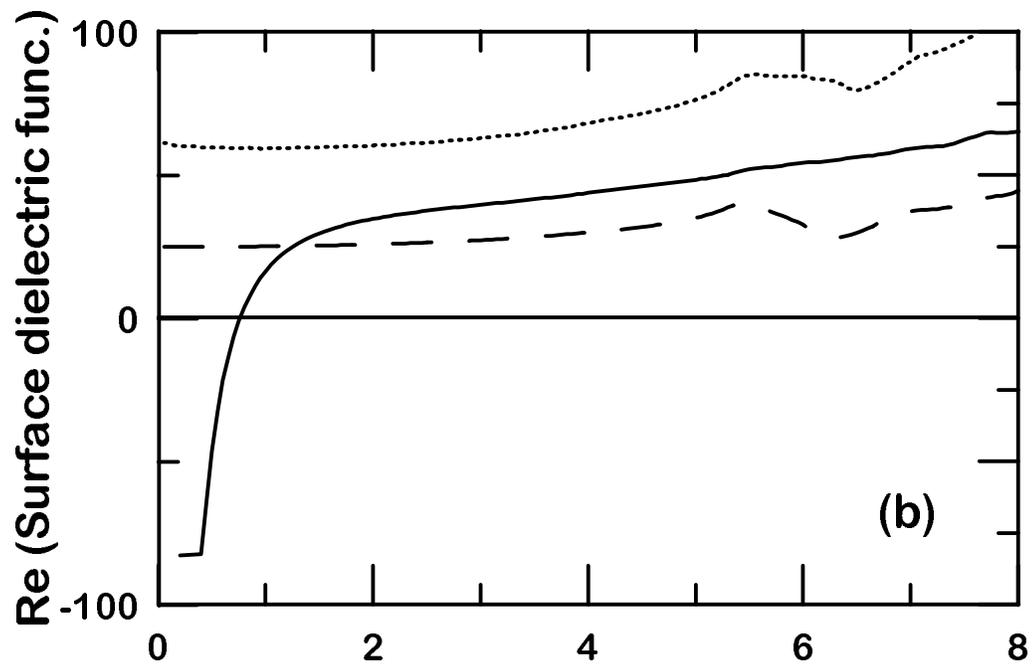
 \begin{figure} 
 \caption {The (a) imaginary part and (b) real part of the {\em
surface} dielectric response. Solid lines correspond to light polarized
along the chains ($x$ axis) $\epsilon_{\rm surf}^{x x}$ , while dotted
lines correspond to light polarized perpendicular to the  chain ($y$
axis), $\epsilon_{\rm surf}^{y y}$. The dashed line corresponds to
light polarized perpendicular to the surface, $\epsilon_{\rm surf}^{z
z}$ .}
 \label{die}
 \end{figure}

In Fig.~\ref{die}, we present the real and imaginary parts of the
surface dielectric tensor $\epsilon_{\rm surf} (\omega)$ calculated
from Eq.~(\ref{df}). The thickness of the surface region used was
$d_{\rm surf} = 2.5$ \AA, which approximately corresponds to two
monolayers (other choices of $d_{\rm surf}$ do not change qualitatively
our results for energies in the bulk gap). The response for light
polarized along the chains ($x$ axis, $\epsilon_{\rm surf}^{x x}$) 
corresponds to the solid lines, while for light polarized in the $y$
direction ($\epsilon_{\rm surf}^{y y}$) is shown by dotted lines, and
the dashed lines correspond to the direction perpendicular to the
surface plane ($\epsilon_{\rm surf}^{z z}$). The imaginary part of
$\epsilon_{\rm surf} (\omega)$ along $x$ shows a strong peak at about
0.1~eV that is 100 times more intense than the rest of the structure
shown in Fig.~\ref{die}. This peak at low-energy is a reflection of the
metallic-like character of the surface along the chains. Then, from 2
to 5.5~eV the dielectric function is nearly constant up to the point when
electron transitions between bulk states become important. 

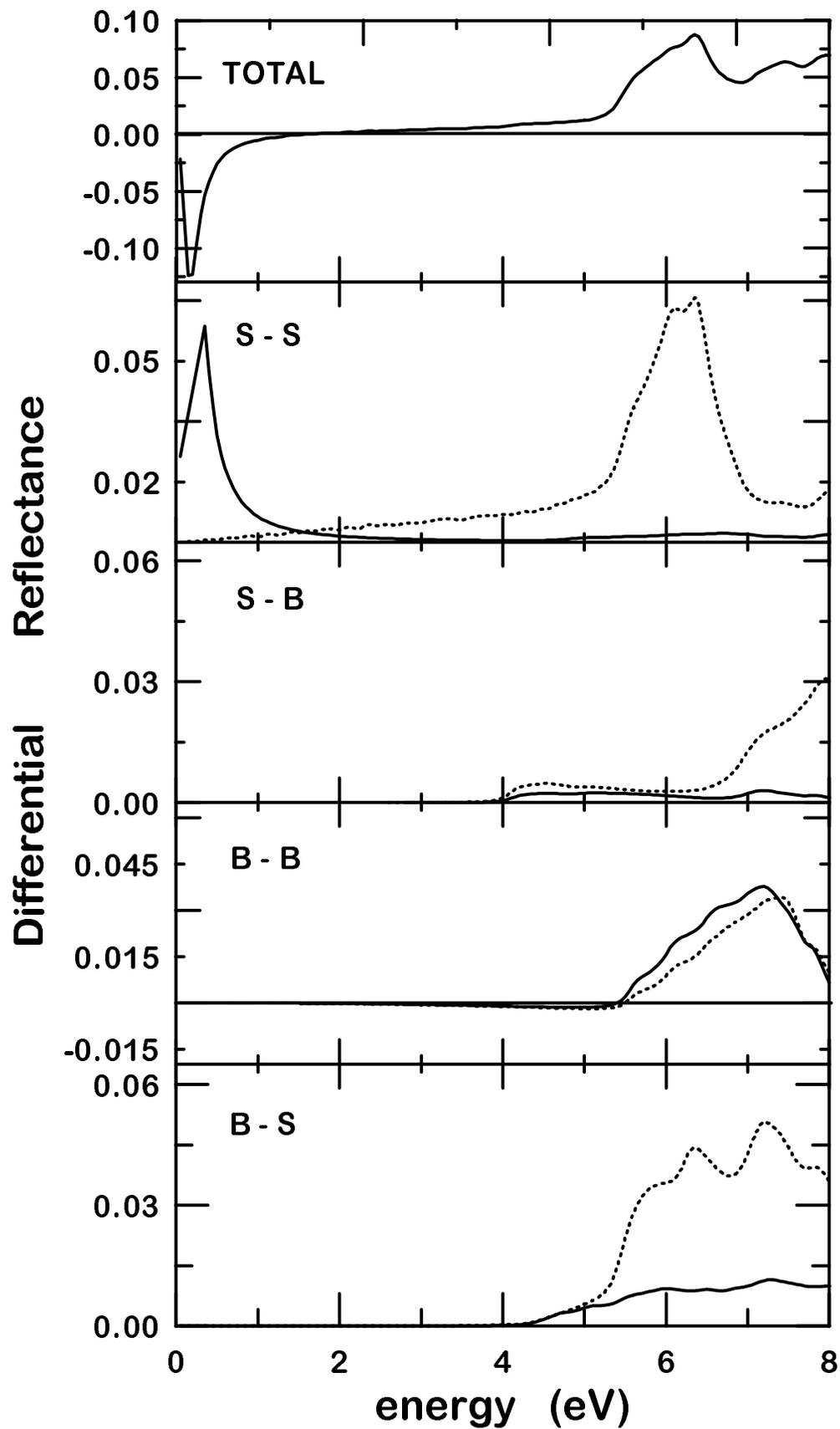
 \begin{figure} 
 \caption {{\em TOTAL} Differential Reflectance spectrum divided into
its {\em S-S}, {\em S-B}, {\em B-B}, and {\em B-S} components. Solid
lines in bottom four panels correspond to light polarized along the
chain ($x$ axis), while the dotted lines correspond to light polarized
perpendicular to  the chains ($y$ axis). }  
 \label{ra}
 \end{figure}

The following discussion about the origin of the main electron
transitions of the surface dielectric function can be seen clearly in
the lower four panels of Fig.\ \ref{ra}, where the reflectance
anisotropy spectrum has been decomposed into the different
contributions, {\em S-S, S-B, B-B, \rm and \em B-S}. For light
polarized in both directions, $x$ and $y$, the dielectric response is
dominated by transitions among surface states ({\em S-S}) up to
$\sim$~4~eV. From about 4~eV the contribution of the transitions from
surface to bulk states ({\em S-B}) and from bulk to surface states
({\em B-S}) becomes important. Note that the gap between the occupied
flat band of surface states along the JK direction on the SBZ (see
Fig.~\ref{sbs}), and the bottom of the bulk conduction band is about
4~eV.  Likewise for the gap between valence bulk states and the
unoccupied surface band at $\Gamma$ point. 
The high density of surface states above and
below the Fermi level along $\Gamma$J results in large {\em S-B} and
{\em B-S} contributions to the Im$\,\epsilon_{\rm surf} (\omega)$ along
the $y$ direction (perpendicular to the chains; shown dotted in Fig.\
2).  Only for this perpendicular direction to the chains, the
Im$\epsilon_{\rm surf} (\omega)$ shows a intense peak centered at about
6~eV due to {\em S-S} transitions. The transitions between bulk states
({\em B-B}) become important from about 5.5~eV onwards, where the
response to $x$ and $y$ polarizations is very similar, as one expects
for cubic semiconductors (notice also the scale change as the traces
are much weaker). For light polarized perpendicular to the surface
plane ($z$ direction), the $\epsilon_{\rm surf}^{z z}$ shows also a
peak around 6~eV mainly due to the first- and second-layer backbond
states (figure not shown). 

In the rest of this section we will discuss the reflectance anisotropy 
and electron energy loss spectra obtained using the calculated surface
dielectric function.

\subsubsection{Reflectance Anisotropy}

In Fig.~\ref{ra}, the top panel shows the differential reflectance
anisotropy spectra for light at normal incidence, $ \left( \frac{
\Delta R}{R_\circ} \right)^y - \left( \frac{ \Delta R}{R_\circ}
\right)^x $, calculated according Eq.~(\ref{dr}), and labelled {\em
TOTAL}.  This has been decomposed in its different contributions, where
the response to light polarized along $x$ (chain axis) corresponds to
the solid line, while the dotted line corresponds to light polarized
along $y$ (perpendicular to the chain).  From the figure, it is clear
that the spectrum shows a large surface anisotropic optical response in
a large range of photon energies. While for  $x$-polarized light the
spectrum shows mainly one peak at low energies, the $y$-polarization
spectrum shows a rich structure for all energies inside the bulk
optical gap.

The intense peak at $\sim$0.1~eV corresponding to $x$-polarized light
is totally determined by transitions between surface states. As we have
explained above, this peak is related with the metallic-like behavior
of the surface along the chain axis. At about 6 eV there are also some
{\em S-S} transitions of weaker intensity for $x$ polarization, and
associated with the resonance states in the conduction band. The rest
of the $x$-polarized spectrum shows a very small contribution from {\em
S-B} and {\em B-S} transitions, compared to the response for
$y$-polarized light.  In fact, the response to light polarized
 perpendicular to the chain axis ($y$ direction) shows much more
structure in a larger energy region within the bulk optical gap. Up to 
$\sim$4~eV the spectrum is only dominated by {\em S-S} transitions. At
4~eV the contribution from {\em S-B} and {\em B-S} transitions starts
and is reflected in the {\em TOTAL} differential spectrum by a
shoulder.  As mentioned above, 4 eV corresponds to the gap between the
flat band surface states around $E_f$ along JK and the bottom of the
conduction band, as well as to the energy difference between valence
bulk states and the unoccupied surface band beginning at 4.5~eV at the
$\Gamma$ point. Then, the intensity enhancement of the {\em S-B}
contribution starting from $\sim$~6.5 eV corresponds to an increase of
the density of the conduction-band states. In all cases, the {\em B-B}
contributions to the {\em TOTAL} differential reflectance spectra in
this range are insignificant, since both polarizations yield nearly
identical contributions.

It is important to notice that this kind of {\em deconvolution} of the
spectrum helps one gain useful insights into the nature of the various
transitions. As we have pointed out, the {\em S-B} contribution starts
at some determined energy ($\approx 4$ eV), as this gap is related with
the conduction band and surface states located around $E_f$. This part
of the spectrum gives then unambiguous information on the position of
the Fermi level with respect to the bulk band structure, and therefore
the energy at which the filled surface states are. Notice that one
important advantage of this optical spectroscopy is the high precision
in measuring the energy at which the electronic transitions occur. The
present results could be important for a future comparison with
reflectance anisotropy measurements in order to better determine the
electronic structure associated to this particular reconstruction of
the surface. We hope this motivates additional experiments.

\subsubsection{EELS}
 
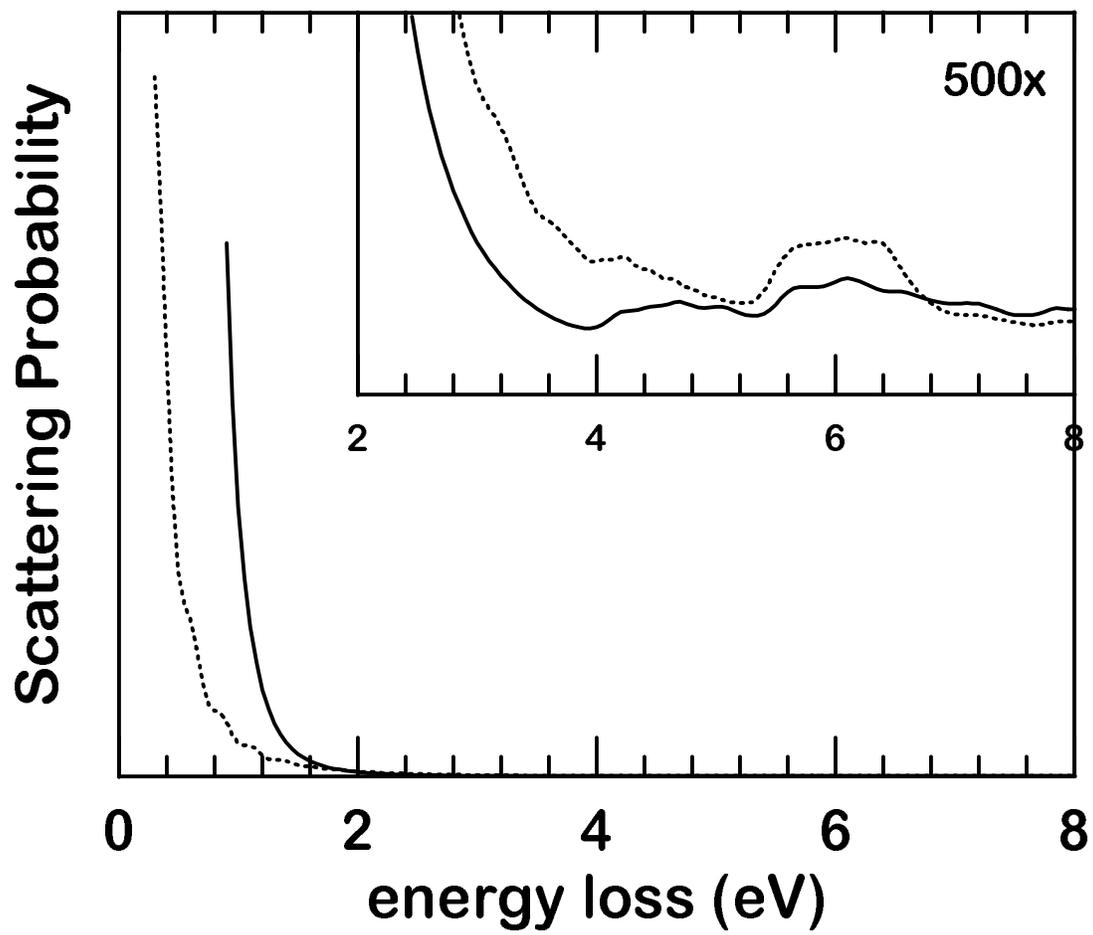
 \begin{figure} 
 \caption {Scattering probability function for electron energy loss,
Eq.\ (4). Solid lines correspond to light polarized along the chain
($x$ axis), while dotted lines correspond to light polarized
perpendicular to the chains ($y$ axis). The inset has been amplified
500 times.}
 \label{eels}
 \end{figure}

The calculated scattering probability of an EELS experiment, using
Eq.~(\ref{pd}), is shown in Fig.~\ref{eels}. The primary energy of the
electron beam was taken equal to 80~eV, with a normal incidence
geometry. The spectrum corresponding to an electron beam polarized
along the chain (solid line) is very different at low energies (less
than 1~eV) than the results for a polarization perpendicular to the
chains (dotted line). The intensity of the $x$-polarized reflected beam
is a few thousand times larger that the beam for $y$ direction
polarization.  At higher energies, from about 4~eV onwards, the two
spectra show similar amplitude and behavior. The inset shows the
scattering probability for energy loss from 2 to 8~eV, where the
intensity has been augmented 500 times. Here, a feature starts at about
4~eV. As we discussed above, these structures are related to the
contribution to the surface dielectric response from {\em S-B} and {\em
B-S} transitions, while the broad peak at about 6 eV is produced by
{\em S-S} transitions.  Notice that the high energy-loss features are
strongly reduced by the decaying prefactor $1/q_{\parallel}$ appearing
in the definition of $P$, Eq.\ (4).

The EELS experiments reported by  Pepper \cite{pepper1} showed a broad
structure centered near 2.1 eV (and width of about 1.7 eV).  The
primary energy of the normal incident electron beam was $E_\circ =
80$~eV. The main spectrum reported by Pepper was obtained by
subtracting the spectra measured for the clean and hydrogenated
surfaces, in order to reduce the effects of a strong elastic peak and
to  enhance the signal due to the reconstruction. The spectra of the
clean and hydrogenated surfaces were obtained from an average over the
SBZ.  In the difference spectrum a minimum gap of about 1~eV was
observed and identified with the effective gap between surface states
at the point J.  The energy resolution of the system is estimated at
0.63~eV, from the width of the elastic peak remnant.  Since the energy
resolution is not optimal in this experiment, it is difficult to make a
direct comparison with theoretical calculations and other experiments.
Moreover, the energy loss measured by this kind of spectroscopy is
generally smaller than the energy difference between occupied and
unoccupied states of the system on its ground state.  Therefore, it is
possible that the observed broad feature at 2.1 eV is related to the
{\em S-S} transitions integrated over the SBZ, and expected to have an
enhanced joint DOS at $\approx 6$ eV\@.  The overall resulting
feature would perhaps be a combination of excitonic downshift and the
high-energy ($1/q_{\parallel}$) suppression factor.  It is clear,
however, that a better-resolution and more detailed EELS study on
this system will be highly desirable.  We will be glad to provide
details of our electronic structure and surface dielectric function
results to interested experimental groups.

\section{Conclusions}

We have investigated the optical response of the C(111)-2$\times$1
surface based on a sp$^3$s$^*$ parameterized tight-binding approach.
The dielectric function of the surface region was calculated and a
large anisotropy was found. This anisotropy of the optical response
is a direct consequence of the surface reconstruction. The dielectric
response of the surface was analyzed in terms of the {\em S-S}, {\em
S-B}, {\em B-B} and {\em B-S} transitions, and important features
corresponding to each type of transitions were found. The reflectance
anisotropy and electron energy loss spectra were calculated in order
to provide direct comparison with experiments.  We can conclude that
these optical spectroscopies combined with theoretical studies, can
help one elucidate the controversial surface electronic structure,
and therefore, the structural and electronic level reconstruction of
this important surface.

\acknowledgements

We are thankful to D. A. Drabold, R. Del Sole, and R. G. Barrera  for
their valuable comments, and to D. R. Alfonso for providing   the
coordinates for the relaxed diamond (111)-2$\times$1 surface.  This
work has been supported in part by the Department of Energy grant no.\
DE--FG02--91ER45334.  C.N. is partly supported by the National 
University of Mexico grants DGAPA-IN-102493 and PADEP-003309. \\

\end{document}